\documentclass[a4paper,11pt]{article}
\usepackage{pos}
\usepackage{graphicx}
\usepackage{wrapfig}

\title{Multi-Messenger High-Energy Results}

\author*[a]{Teresa Montaruli}

\affiliation[a]{Département de Physique Nucléaire et Corpusculaire, Faculté de Sciences, Université de Genève, 1205 Genève}

\emailAdd{teresa.montaruli@unige.ch}
\abstract{Multi-messenger high-energy astrophysics is the extension of the multi-wavelength exploration of the cosmos with multiple messengers with a common origin, including neutrinos, gravitational waves, and cosmic rays. This branch of astrophysics has currently achieved the potential to unravel the origin of cosmic rays and how sources accelerate them, their relation to the diffuse radiation in the extra-galactic space, and their role to forge their galaxies of origin while they wander in their magnetic fields for millions of years. 

Neutrino astronomy produced its major scientific milestone with the discovery by IceCube of a diffuse flux at energies above 60~TeV with intensity comparable to a predicted upper limit to the flux from extra-galactic sources of ultra-high energy cosmic rays. More recent results provide the first strong evidence of a standalone neutrino source and a highly probable coincidence of a neutrino alert with gamma rays. These results of IceCube indicate that neutrino astronomy can complement photon astronomy also providing insights into opaque sources of high-energy radiation. Starburst galaxies and jetted black holes in active galaxies are favored candidates to explain the diffuse cosmic neutrino background at $> 60$~TeV energies and its relation to the extragalactic background light. Additionally, gamma-ray bursts remain an intriguing mystery now enriched by joint observations of gamma-rays and gravitational waves. Ground-based detection of gamma-ray burst emissions with energies up to more than 10~TeV challenges the standard fireball model as well as the non-observation of neutrinos.

The galactic diffuse flux, produced by cosmic ray interactions on the interstellar matter of our galaxy and peaking at lower energies, is within the reach of neutrino detectors. Together with the measured galactic gamma-ray flux up to PeV energies, they will shed light on the knee region of cosmic rays and the possible existence of dark matter in the Galactic plane. 

In the future, more work will be done in IceCube and deep sea and lake neutrino telescopes to use further low-energy cascades for cosmic source searches thanks to improved descriptions of detection media and deep learning methods.

These aspects were discussed at the conference and are summarised in this write-up, and when necessary more recent results will be referred to.
}

\FullConference{%
  7th Heidelberg International Symposium on High-Energy Gamma-Ray Astronomy (Gamma2022)\\
  4-8 July 2022\\
  Barcelona, Spain\\}

\begin{document}
\maketitle

\section{The Gamma-ray messengers}
\label{sec1}

Cosmic Rays (CR) constitute the ``astroparticle'' component of cosmic radiation, composed as well of photons from radio to gamma-ray energies. 
The discovery of the CR flux, often dated in 1912 and attributed to V. Hess, finds its root in the work of many scientists already from the previous century with the discovery of radioactivity in 1986 by Becquerel. 

In 1953, indirect detection of gamma rays from the ground was at its primitive attempts with a photomultiplier (PMT) and a mirror in a garbage can by Galbraith and Jelley~\cite{galbraith,1996SSRv...75....1W}. 
Their observations proved right the prediction of P.M.S. Blackett that Cherenkov radiation emitted by high-energy CRs should contribute to the light in the night sky~\cite{blackett}. 
Blackett had also estimated that the Cherenkov radiation produced by CRs traversing the atmosphere comprised $\sim 10^{-4}$ of the night-sky background, and commented that `{\em presumably such a small intensity of light could not be detected by normal methods}'.

\begin{wrapfigure}{L}{0.5\textwidth}
  \centering     \includegraphics[width=0.5\textwidth]{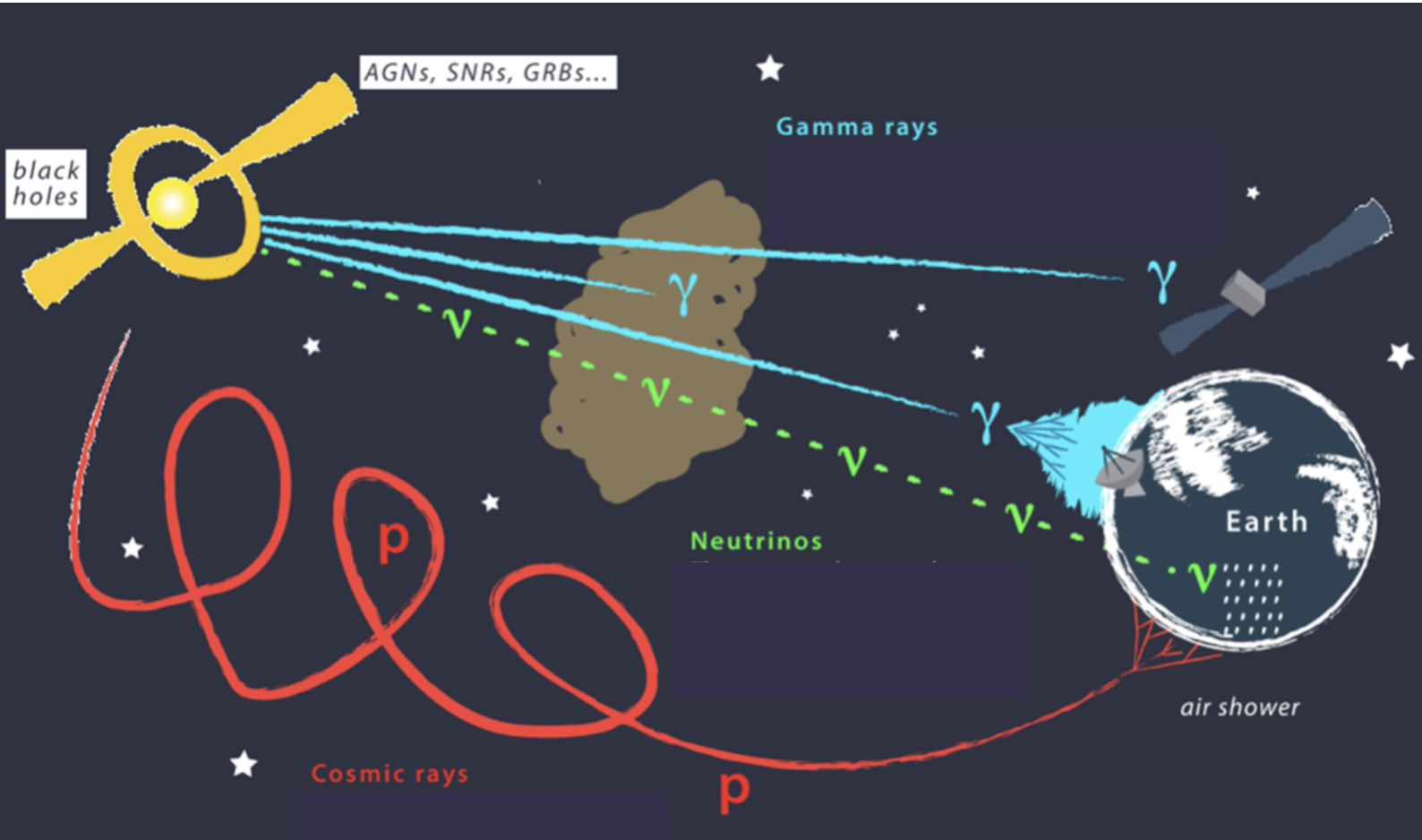} 
    \caption{\label{fig:MM} \small \it Multi-messengers from hadronic accelerators (credit J.A. Aguilar \& J. Yang).} 
\vspace{-15pt}
\end{wrapfigure}

The first detection of a TeV source from the ground had to wait until 1988 when the Whipple Imaging Air Cherenkov Telescope (IACT) in Arizona imaged the Crab Nebula~\cite{1989ApJ...342..379W}. This source is now known as the 'standard candle' of gamma-ray astronomy as most telescopes by now cross-calibrate their data against its precisely measured spectrum. Nonetheless, its pulsar embedded in the supernova remnant (SNR) offered many unpredictable surprises \cite{Buhler:2013zrp}, such as GeV flares of the order of days~\cite{2011Sci...331..739A}, emissions of pulsations from the pulsar to $\sim 1$~TeV~\cite{2016A&A...585A.133A}, and photons up to PeV energy probably from the shock in the nebula~\cite{LHAASO:2021cbz}. The morphology of the source to the level of imaging the pulsar wind nebula (PWN) was reconstructed by H.E.S.S. thanks to the large statistics accumulated in 22~h of observations during 6~yr up to an angular resolution of about $0.14^\circ$ above 300 GeV~\cite{HESS:2019beq}. The source extension of the gamma-ray emission is smaller than that in the ultraviolet band and significantly larger than in the hard X-rays seen by Chandra. The Synchrotron Self Compton model (SSC) explains observations and data favor a wind solid angle of 6.5~sr and a termination shock radius of 0.13~pc, confirming Chandra's results. 

Gamma-ray astrophysics has made huge progress in recent years, as witnessed by the huge scientific impact of the Fermi-LAT and GBM experiments launched in 2008. Fermi-LAT published its 12~yr catalog 4FGL dominated by blazars (about 62\% on a sample of 5064 sources, while 239 are identified pulsars)~\cite{Fermi-LAT:2022byn}. In the future, the gamma-ray search from space is challenged by the absence of a mission able to cover the gap between 100~keV and 1~GeV, as highlighted in the Astro2020 Decadal Survey document and in the dedicated Snowmass study \cite{Engel:2022bgx}. The AMEGO-X~\cite{Caputo:2022xpx} and e-ASTROGAM \cite{e-ASTROGAM:2017pxr} projects have been proposed. 

\begin{wrapfigure}{L}{0.5\textwidth}
  \centering     \includegraphics[width=0.5\textwidth]{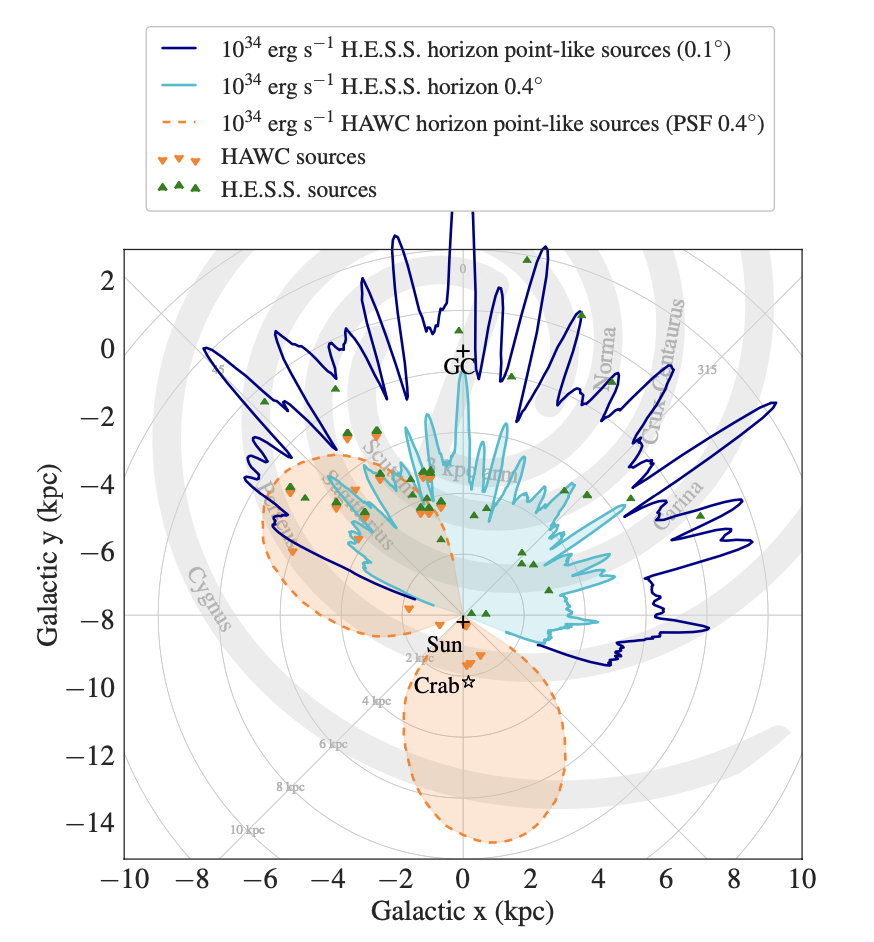} 
\caption{\label{fig:HESS-HAWC} 
View of the Galaxy with schematic spiral arms in gray. The H.E.S.S. horizon at $b = 0^\circ$ for a $5\sigma$ detection of a point-like (extended) source is in dark (light) blue. The HAWC horizon is in orange. The sources located further than the horizon are brighter than $10^{34}$ erg s$^{-1}$. The sources with known distance and detected by H.E.S.S. and HAWC are the green and orange triangles.  From~\cite{2021ApJ...917....6A}.} 
\vspace{-10pt}
\end{wrapfigure}

The current generation of IACTs has reached maturity to the level of scanning the Galactic Plane up to 10-20 mCrab \cite{2018A&A...612A...1H} and pinning down the morphology of sources. Currently, the most extensive catalog of TeV sources, the TeVCat, lists 252  sources in Jan.~2023 \cite{tevcat}. Many of them have been detected by arrays of a few IACTs with mirrors of mirror diameters between 12 and 24~m: H.E.S.S.~\cite{HESSweb}, MAGIC~\cite{MAGICweb} and VERITAS~\cite{VERITASweb}. With the advent of the Cherenkov Telescope array sensitivity to the Galactic Plane scan will improve by one order of magnitude as well as the angular precision. The energy threshold will reach down to $10-20$~GeV \cite{CTAConsortium:2018tzg,CTAweb}.

Extensive air shower arrays (EAS) and hybrid arrays complement IACTs thanks to their duty cycle which is larger than 90\% and their large field of view (FoV) of $\sim 2$~sr. IACTs are limited to FoV of few degrees and to duty cycles of the order of 10\%, but they achieve better angular and energy resolution than EAS, and reach down to a lower energy threshold, where the gamma-ray horizon is larger, source fluxes are higher and overlapping with direct measurements on satellite.
HAWC~\cite{HAWCweb} is an array of 300 water Cherenkov detectors (WCDs) located in Mexico close to the Pico de Orizaba, at an altitude of 4100 m. It has a  collection area of $\sim {10}^{4}\,{{\rm{m}}}^{2}$ above 500 GeV, close to the geometrical area.
H.E.S.S. and HAWC complement each other beyond 1~TeV for the Galactic Plane sources, depending on their extension \cite{2021ApJ...917....6A}. For searches of sources with extension of $0.4^\circ$ and more, H.E.S.S. loses sensitivity as IACTs have limited FoV (see Fig.~\ref{fig:HESS-HAWC}). The shown sensitivity includes 2700 hr of selected data, with 78 detected gamma-ray sources selected \cite{2018A&A...612A...1H} and 1523 d of HAWC data. 
The precursor of HAWC, Milagro, pioneered the water Cherenkov detection with muon discrimination in a pond, now adopted by LHAASO in China at a much larger scale with a 78'000~ton of water pond at 4410~m altitude\cite{LHAASOweb}. LHAASO proved its reach to PeV gamma-ray astronomy for the Galactic Plane \cite{LHAASO:2021cbz}, opening a new frontier that will tackle issues as dark matter and the CR knee (see Sec.~\ref{sec:grantedfluxes}).

The complementarity between IACTs and EAS is of relevance for the field of ground-based astronomy: the future CTA Observatory (CTAO) \cite{CTAweb,CTAConsortium:2018tzg}, currently under construction, will operate together with the LHAASO array and eventually with SWGO \cite{Hinton:2021rvp}, now at the proposal stage. This will guarantee continuous observation of transient sources as well exploration of extended regions.  

\section{Neutrino messengers and their telescopes}
\label{sec:detectors}

In 1960, Greisen suspected that `{\it within the next decade cosmic ray neutrino detection will become one of the tools of both physics and neutrino astronomy}’~\cite{Greisen:1960wc}. He also envisaged neutrino astronomy's connection with CRs and with the emerging field of gamma-ray astronomy. This connection is artistically shown in Fig.~\ref{fig:MM}, where a CR accelerator produces neutral messengers, neutrinos, and gamma-rays, in interactions with the source radiation or matter. Gamma rays are reprocessed on the way to the Earth or satellite-based detectors, lose energy and are absorbed due to interactions on radiation backgrounds pervading the universe, while neutrinos travel undisturbed. Neutrinos can extend the accessible horizon of gamma rays to regions where they are absorbed.
Despite their different cross-sections, and so their different horizons and propagation properties, the coincident observation of neutrinos and gamma rays from a source is a strong indication of the acceleration of protons/nuclei.
Additionally, gamma-ray and neutrino messengers preserve the direction of their sources, while CRs, are subject to deflections that randomize their paths from the source due to intergalactic and galactic magnetic fields.

\begin{wrapfigure}{L}{0.55\textwidth} 
  \centering
      \includegraphics[width=0.55\textwidth]{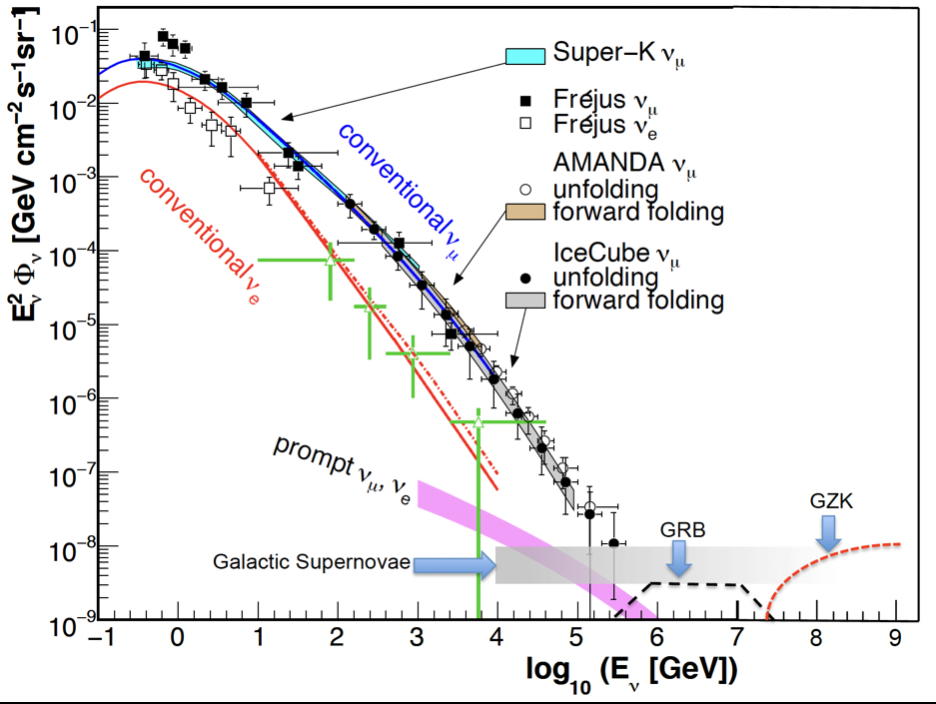} 
    \caption{\label{fig:atmo} \small \it
      The measured atmospheric $\nu$ flux by IceCube for the muon and electron components. The prompt and cosmic $\nu$ fluxes are indicated for illustration.
      } 
\vspace{-10pt}
\end{wrapfigure}

In 1960, Markov discussed in a proceeding his vision of deep natural media used as neutrino detectors~\cite{Markov:1960vja}, and in 1965 Reines and Cowans pioneered Cherenkov detection of anti-electron neutrinos through inverse-beta decay in the East Rand gold mine in South Africa at 3.5~km depth~\cite{Cowan:1956rrn}. 
After 2 decades, the optical observation of the cataclysmic core-collapse supernova 1987A followed a few hours after the detection of a few neutrino events by the underground detectors Kamiokande~\cite{Hirata:1988ad}, IMB~\cite{IMB:1987klg} and Baksan~\cite{Alekseev:1988gp}. The 2002 Nobel prize was awarded to M. Koshiba for his leadership of the Kamiokande collaboration to achieve this first milestone of multi-messenger astrophysics. The future observation of a similar event would allow a revolution in understanding the formation of compact objects after a supernova collapse by using neutrinos, gravitational waves (GWs), and the electromagnetic spectrum~\cite{Nakamura:2016kkl}.

\begin{wrapfigure}{L}{0.5\textwidth} 
  \centering
      \includegraphics[width=0.5\textwidth]{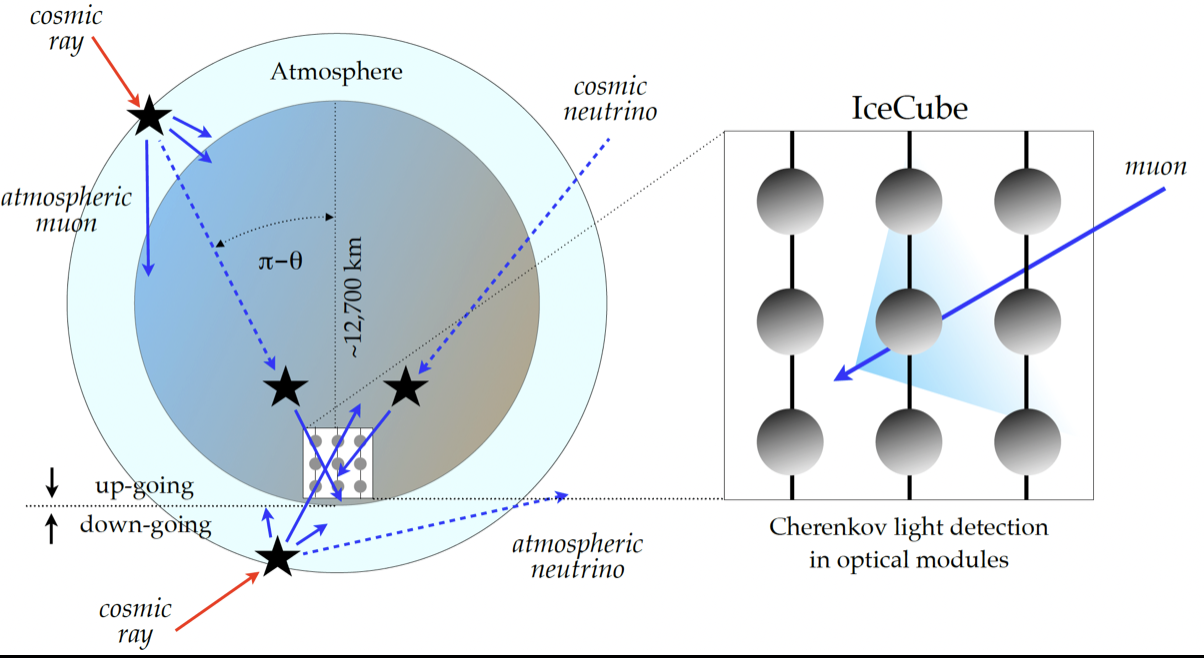} 
    \caption{\label{fig:principle} \small \it Principle of indirect $\nu$ detection in NTs.
      } 
\vspace{-10pt}
\end{wrapfigure}

Since Markov's proposal, neutrino telescopes (NTs) reached their maturity at $km^3$-scale with IceCube in operation. KM3NeT \cite{KM3NeT:2022pnv} and GVD \cite{Baikal:2022chp}, now in construction, already take data and are confirming the discovered IceCube diffuse flux of astrophysical neutrinos by IceCube (see 
Fig.~\ref{fig:diffuse_MM} discussed later).

Neutrinos might come directly from the cosmos or they are generated from CR interactions in the atmosphere that produce also muons, which above TeV energies are capable of penetrating to the detector depth and are screened from very horizontal and upgoing direction by the earth
(see Fig.~\ref{fig:principle}). 
Typical rates in IceCube are 2.5~kHz for downgoing atmospheric muons, about $2 \times 10^5$ upgoing atmospheric $\nu$s/yr, and above 60~TeV  $\sim 10~\nu$s/yr, mostly of cosmic origin.

NTs and IACTs share the same feature of detecting messengers indirectly and inferring their properties, e.g. direction and energy, through Monte Carlo simulations. These simulate their flux, interactions of particles and their propagation in well-detailed media with their inhomogeneities and detector response. The systematic errors of measurements are related to how accurately these aspects are modeled. Huge simulated event statistics are required corresponding to short equivalent times with respect to the total acquisition time as the dimension of the detector effective areas are big. Additionally, in IACTs and NTs the atmospheric background also includes components identical to the signal. For IACTs, electrons developed in the EAS are an irreducible background for photons and for cosmic neutrinos they are atmospheric neutrinos (see Fig.~\ref{fig:atmo}). 

\begin{wrapfigure}{L}{0.55\textwidth}
  \centering
      \includegraphics[width=0.55\textwidth]{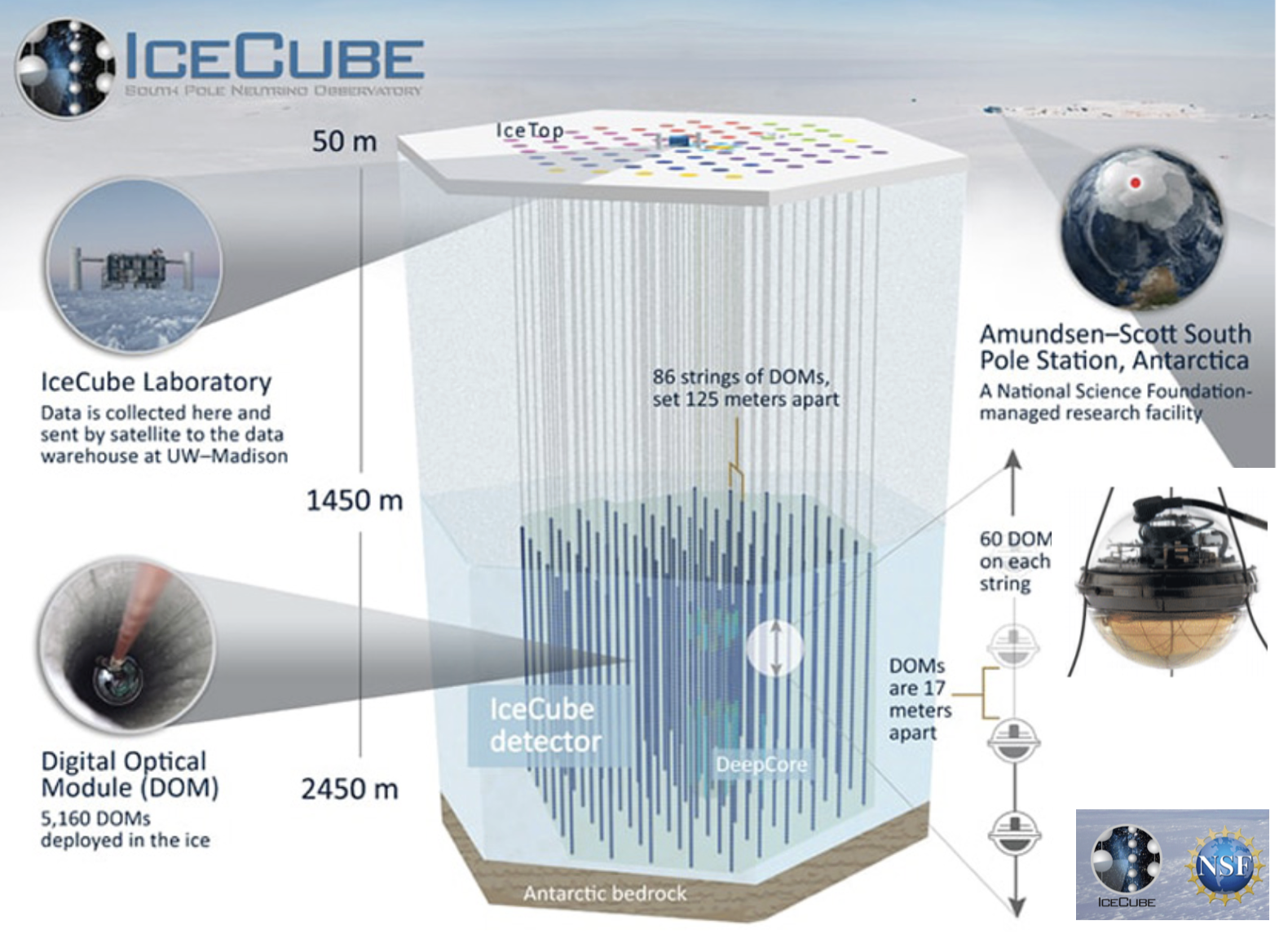} 
    \caption{\label{fig:NT} \small \it The IceCube in-ice detector, the IceTop surface array on top of the strings and a downward-looking PMT.
      } 
\vspace{-10pt}
\end{wrapfigure}

IceCube, located at the South Pole, is an array of 5160 digital optical modules (DOMs) along 86 strings, synchronized to ns level instrumenting the deep ice between about 1450~m and 2450~m. Strings are separated by about 125~m and with a vertical distance between DOMs of about 17~m (see Fig.~\ref{fig:NT}).
The DOMs measure the time and energy of photons emitted by the charged secondaries of neutrinos through Cherenkov light emission with an angle of about $\sim 41^{\circ}$ (see Fig.~\ref{fig:principle}).  
Among the 86 strings, 8 are more densely instrumented and named DeepCore for neutrino oscillations and dark matter studies below 100~GeV.  
A random background of $\sim 1$~kHz is emitted by relativistic positrons in the $\beta$-decay of $^{40}K$ from the 0.5"-thick spherical glass of the DOMs~\cite{IceCube:2016zyt}. 
The time resolution of PMTs is $\sim 2$~ns, when illuminated by narrow pulses, but the dispersion of photons due to scattering in the ice is the limiting factor for the reconstruction~\cite{IceCube:2010dpc}. 

Depending on the neutrino flavor and so on the different secondaries they produce in their interactions, there are two main topologies of neutrino events: tracks from muon neutrino deep inelastic charge current interactions and cascades from other flavor neutrinos and neutral current interactions of all flavor neutrinos. Hints for ``double bump events'', characteristic of tau neutrinos above few PeV have been recently found \cite{IceCube:2015vkp}. As tau neutrino production in the atmosphere has a low probability at the observed high energies, they can be tracers of neutrino oscillations along cosmic baselines. Additionally, one cascade event has been measured possibly generated by an anti-electron neutrino event of about 6~PeV \cite{IceCube:2021rpz} in the Glashow resonance region.

Recently, the modeling of the light propagation properties of the ice has been extremely improved, with the addition of an asymmetric light diffusion effect in the birefringent poly-cristalline micro-structure of the ice~\cite{2021JInst..16C9014R} enabling a better angular resolution also for the cascade channel. Reconstruction techniques are also advancing by deep learning employment, such as Graphic Neural Networks \cite{IceCube:2022njh} and Convolutional Neural Networks \cite{Abbasi:2021ryj} running on GPUs with enormous advantages for processing time. 
As it is understandable from Fig.~\ref{fig:atmo}, improved samples of cascades, dominated by electron neutrinos, profit of the lower background of atmospheric neutrinos with respect to muon neutrinos. The typical angular resolution of cascade samples is now at the level of $5^{\circ}$ in ice, comparable to that of sea detectors. Nonetheless, the muon channel remains the only one with $\sim 0.2^{\circ}$ angular resolution above 10~TeV \cite{IceCube:2019cia}. 

\begin{wrapfigure}{L}{0.5\textwidth} 
  \centering
      \includegraphics[width=0.5\textwidth]{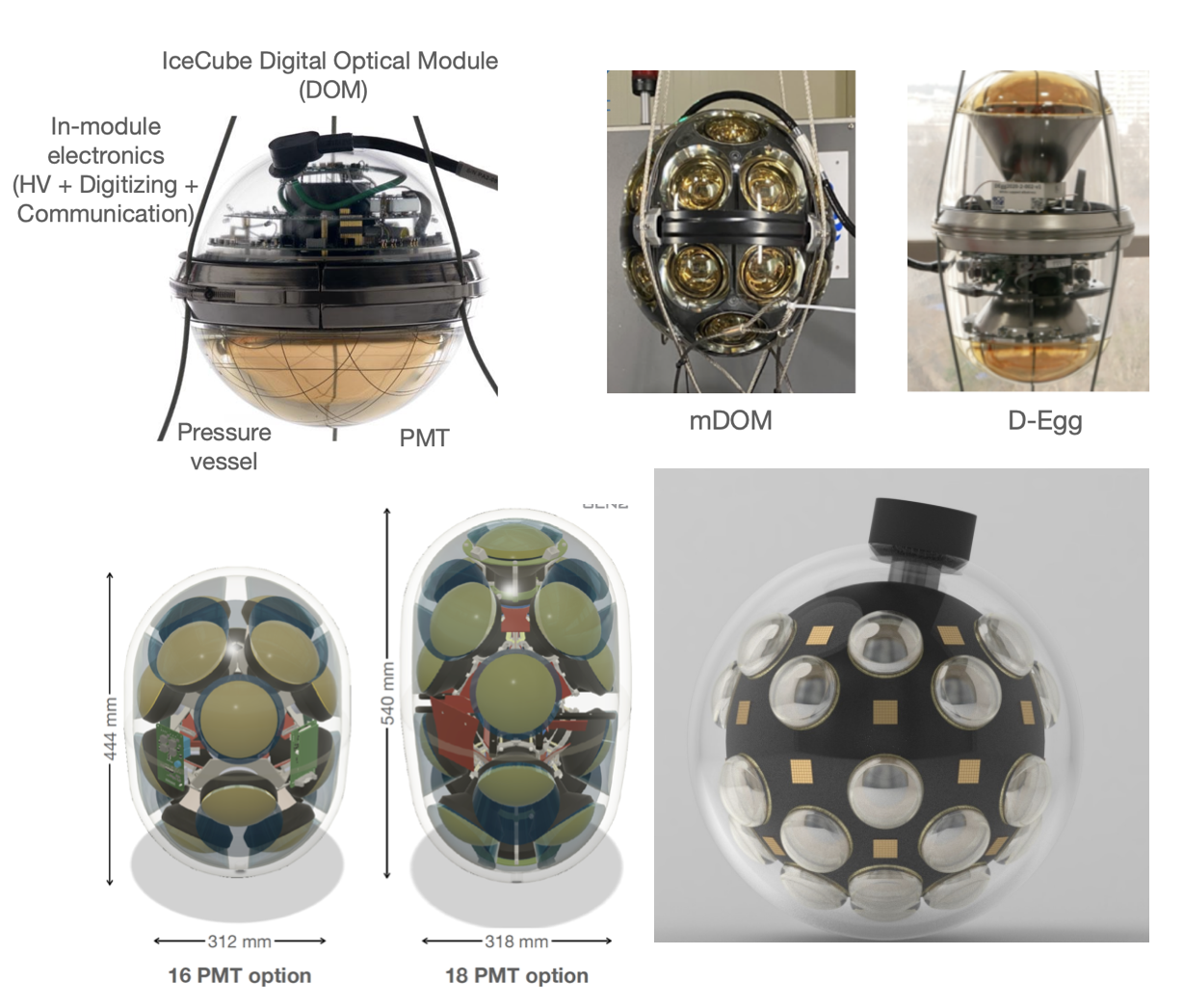} 
    \caption{\label{fig:Photosensors} \small \it
      IceCube DOM, D-Egg and mDOM prototypes for the IceCube Upgrade and the designs for IceCube-Gen2, based on PMTs; the TRIDENT concept~\cite{Ye:2022vbk}.} 
\vspace{-10pt}
\end{wrapfigure}

The future of IceCube has started: the upgrade of IceCube, namely the addition of 7 dense strings with 700 optical modules followed by IceCube-Gen2. Its science scope, described in \cite{2021JPhG...48f0501A}, expands that of IceCube.
The evolution of the optical modules is shown in Fig.~\ref{fig:Photosensors}, where the compromise proposal for IceCube-Gen2, resulting from the experience based on two developed prototypes, is shown. The design inherits from the KM3NeT concept \cite{KM3NeT:2022pnv} and profits of better uniformity and smaller transit time spread of small PMTs compared to larger ones. 

Traditionally, NTs and many other Cherenkov detectors in astroparticle are based on very advanced PMT technology. 
Recently, gamma-ray astronomy has pioneered also SiPMs \cite{FACT,Heller2017,Montaruli_icrc}, which might become an option for the future of NTs. A recently proposed concept for a neutrino telescope design in the Pacific is also shown in the picture, combining the large surface coverage of PMTs with the fast rise time of SiPMs~\cite{2022icrc.confE1043H,Ye:2022vbk}. Simulations indicate an improvement of about 40\% in the angular resolution.


\section{Extra-galactic diffuse background radiation and cosmic particle fluxes}
\label{sec3}

Nowadays the most compelling aspects of astroparticle physics concern the unknown origin of CRs, how they emerge out of their powerful sources and how acceleration mechanisms work and evolve with time, CR diffusion and influence on the evolution of their hosting galaxies, their probing potential of the magnetic fields and of the dark matter in the universe. 

In 1990, Ressel and Turner presented the Grand Unified Photon Spectrum (GUPS)~\cite{1990ComAp..14..323R}. On the extragalactic background radiation or light (EBL) spanning about 19 orders of magnitude in energy from the radio to the gamma-ray band, they superimposed the CR flux. This extends from the energy of the mass of protons to beyond $10^{20}$~eV, hinting to the existence of extreme extragalactic accelerators beyond the reach of IACTs. 
In fact, high-energy photons are absorbed during propagation by pair production on the diffuse radiation limiting the gamma-ray horizon. 

Based on the observed energy density of $3 \times 10^{-19}$~eV/cm$^3$, calculated integrating the CR flux above $10^{17}$~eV~\footnote{The CR flux is related to the energy density of their sources by $\rho_E = 4 \pi \int \frac{E}{\beta c} \frac{dN_{CR}}{dE} dE$. As a comparison to the extragalactic region energy density in the text, if we integrate between $1-10^6$~GeV, we obtain the energy density of galactic CRs of $\sim 1$~eV/cm$^3$, comparable to the galactic magnetic field energy density.}, it was speculated that gamma-ray bursts (GRBs) and black holes and their jets embedded in active galactic nuclei (AGNs) can sustain ultra-high energy CRs (UHECRs)~ \cite{1998PhRvD..59b3002W}. 

\begin{wrapfigure}{L}{0.65\textwidth} 
  \centering
      \includegraphics[width=0.65\textwidth]{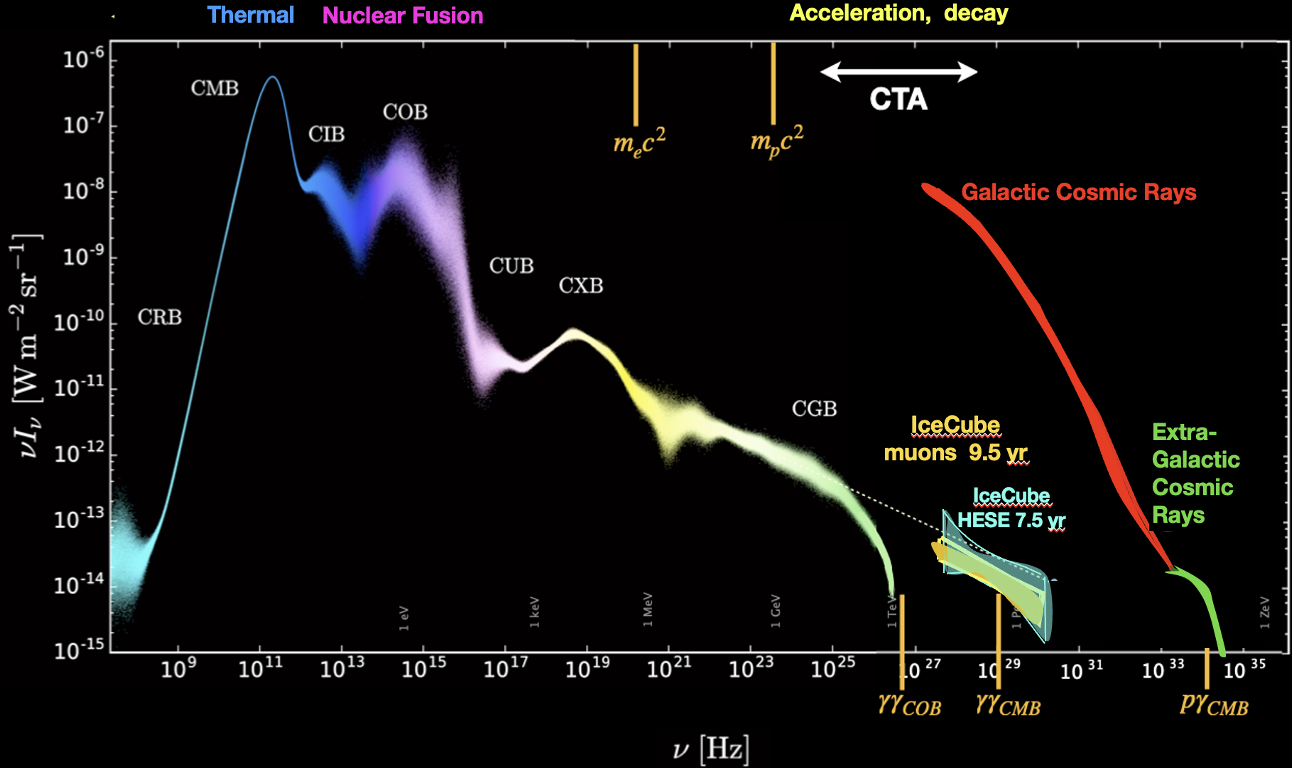} 
    \caption{\label{fig:diffuse_MM} \small \it EBL energy spectrum from radio to gamma rays \cite{Hill:2018trh,DeAngelis:2018lra}. The CR spectrum is indicatively drawn from $10^{13}$~eV~\cite{ParticleDataGroup:2022pth}. The diffuse spectra of the IceCube HESE sample in 7.5~yr~\cite{IceCube:2020wum} and the diffuse muon tracks accumulated in 9.5~yr~\cite{IceCube:2021uhz} are shown.} 
\vspace{-10pt}
\end{wrapfigure}

Since the first GUPS was presented, the EBL spectrum has been further updated (see the colored spectral emission in Fig.~\ref{fig:diffuse_MM} from \cite{Hill:2018trh,DeAngelis:2018lra}) using measurements of many experiments. The EBL is dominated by the thermal relic radiation from the last scattering surface observed today, the cosmic microwave background (CMB) of $\sim 400$~photons/cm$^3$. At about 10\% of it, the Cosmic Optical Background (COB) emission from stars and the Cosmic Infrared Background (CIB), namely the optical radiation reprocessed by dust and attenuated by it to be re-radiated in the infrared band, allow understanding of star formation as diagnostic for stellar nucleosynthesis, mass accretion onto black hole processes and gravitational collapse of stars. 

 The COB and CIB bumps in the EBL are studied by gamma-ray telescopes in space and on the ground as they can sample a large population of flaring AGNs at different redshifts z. They infer their injection spectra from the measured ones at energies where the EBL attenuation is negligible and compare them to higher energy attenuated spectra to estimate the optical depth as a function of energy and redshift~\cite{Biteau:2015xpa,Fermi-LAT:2018lqt,MAGIC:2019ozu}
These measurements also offer a new approach to infer the Hubble constant of the late universe and to constrain the matter content in the universe, though these estimates depend on the still large 
uncertainty on the number density of the EBL as a function of energy~\cite{2019ApJ...885..137D}. 
CTAO, thanks to three different sizes of telescopes, will improve this measurement reaching redshifts up to $z \sim 2$ thanks to the better sensitivity by about a factor of 10 and its wide energy range from $\sim 20$~GeV to 300~TeV \cite{CTA:2020hii}. 

\begin{wrapfigure}{L}{0.65\textwidth} 
  \centering
      \includegraphics[width=0.65\textwidth]{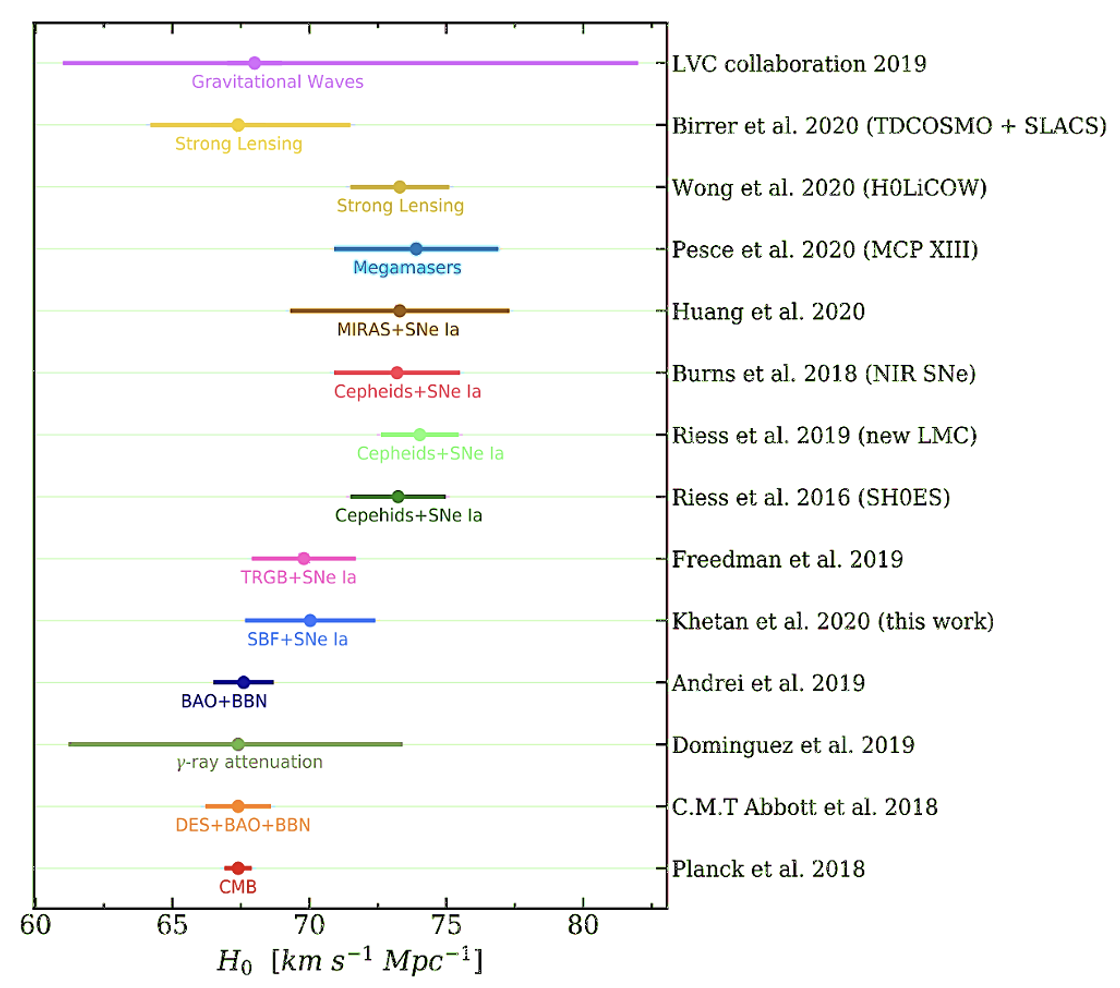} 
    \caption{\label{fig:Hubble} \small \it Compilation of Hubble constant late and early universe measurements with inputs from multi-messengers (gamma-rays and GWs). From~\cite{Khetan:2020hmh}.} 
\vspace{-10pt}
\end{wrapfigure}

Together with gamma rays, another new messenger, GWs, contributed such a measurement (see Fig.~\ref{fig:Hubble} from Ref.~\cite{Khetan:2020hmh}).
Unlike gamma-rays, standard sirens, such as the famous GW170817 neutron star merger, provide the absolute luminous distance $D_L$ from the fit of the gravitational chirp data, hence a simple relation provides $H_0$ from the known redshift from electromagnetic counterparts: $H_0 D_L = c z$ \cite{Abbott:2017xzu}.
The  uncertainty on $H_0$ principally depends on the degeneracy between distance and inclination of the plane of the binary system. Both gamma-ray telescopes and GW interferometers have still a limited horizon, despite both GWs and neutrinos potentially cover a horizon reaching the early universe and propagating through absorbing media in cosmic sources. 
Nonetheless, future advanced detectors may make them players to solve the controversy on the early and late universe measurements of the Hubble constant \cite{Chen:2017rfc}.


At higher energies than the region of the COB and CIB in Fig.~\ref{fig:diffuse_MM}, the gamma-ray part of the EBL is named Extragalactic Gamma-ray Background (EGB) and it has been measured by Fermi-LAT \cite{Ackermann2016}. When the Galactic Plane contribution is subtracted, only the extragalactic diffuse emission from faint and unresolved extragalactic sources remains, mostly blazars and starburst galaxies. The EGB gives the non-thermal perspective on the cosmos, together with the extragalactic CR flux and the diffuse cosmic flux of neutrinos recently discovered by the IceCube km$^3$ neutrino telescope in Fig.~\ref{fig:diffuse_MM} \cite{Aartsen2013,1989ApJ...342..379W,IceCube:2021uhz}. These events are cascade-like when induced by electron or tau neutrinos, or neutral current interactions of all flavor neutrinos with their vertex inside a detector fiducial volume. They are called high-energy starting events (HESE) and selected at an energy beyond 60~TeV and over 7.5 yr of data taking \cite{IceCube:2020wum} is dominated by down-going cascade-like events with a limited angular resolution of $\sim 10^{\circ}$. Another sample is up-going-muons induced by muon neutrinos selected in an independent analysis using 9.5~yr of data taking \cite{IceCube:2021uhz}. Both the HESE and muon track samples constitute evidence with larger significance than $5\sigma$ that cosmic neutrinos are required on top of the background of atmospheric muon and neutrinos to explain the IceCube data. 

As seen from Fig.~\ref{fig:diffuse_MM}, the highest energy end of the multi-messenger plot reveals comparable energy rate density for the UHECRs measured by the Pierre Auger observatory (PAO) \cite{Abreu_2021,Ivanov:2021mkn} and the IceCube neutrinos between 60~TeV and PeV energies. This can be explained by a unified origin (as already hypothesized by Ressel and Turner) assuming photo-meson interactions in extragalactic sources~\cite{2022arXiv220200694H}. 
The ‘UHECR-neutrino unification’ was discussed in detail by Waxman and Bahcall in 1998.
The Waxman and Bahcall upper bound or {\bf calorimetric limit}~\footnote{The upper bound was criticized for having been calculated under strong assumptions \cite{Mannheim:1998wp}, namely that UHECRs are dominantly protons, while in the highest energy end of the spectrum, PAO measures heavier composition. The assumed spectrum was $E^{-2}$ fitted from above $10^{19}$~eV and extrapolated to lower energies where the energy spectrum might be different.} is obtained for a fully efficient system for CR energy loss into pion production. The chain of pion decay relates neutrinos and gamma-ray secondaries to primary protons or nuclei, namely CRs, as they are the results of proton–proton or proton–photon interactions in cosmic ray sources accelerating protons and ionized nuclei. The calorietric condition corresponds to a diffuse extragalactic neutrino flux upper limit of about $E^2 dN/dE \sim 10^{-8}$~GeV s$^{-1}$ sr$^{-1}$ cm$^{-2}$. Below this boundary, the system is 'optically thin' and implies that both UHECR and neutrinos originate from systems with an optical depth of less than $\tau_{p\gamma}\sim 0.6$ \cite{PhysRevD.102.083023}.

Higher neutrino fluxes than this upper bound can be produced in hidden-core AGNs or opaque sources from which only neutrinos escape. The IceCube diffuse flux order of magnitude seems to indicate a large contribution from this topology of sources. Additionally, a joint study between neutrino telescopes (IceCube and ANTARES) and UHECR experiments (PAO and Telescope Array) excluded possible correlations between UHECR directions and neutrinos \cite{IceCube:2022osb}. This could be due to opaque sources contributing to the diffuse neutrino flux but also to the different horizons of the two cosmic messengers~\footnote{As a matter of fact, the highest energy neutrinos observed by IceCube could be dominated by sources beyond the $\mathcal{O}(100)$~Mpc limited horizon of UHECRs.}. 
It should be noticed that, if the UHECR composition is dominated by heavier nuclei than protons, the predicted neutrino flux might be difficult to reach for current detectors \cite{Heinze:2019jou}. 
Hence, it remains open the question:
{\em Which are the sources contributing to the measured diffuse IceCube neutrino flux?}.

 \subsection{Gamma-ray Bursts}
 
In the `canonical’ single-zone standard model of GRBs (see \cite{1995PASP..107..803U} and references therein), the prompt phase of gamma-ray emission is due to the ejecta forming an expanding fireball under thermal pressure, with the efficient conversion of thermal energy to kinetic energy, then becoming optically thin. These ejecta are caused by the collapse of a rapidly spinning massive star or a binary neutron star merger event forming a powerful engine launching relativistic jets. These dissipate their kinetic energy through accelerating shocks of electrons and protons formed by jet collisions. The prompt phase is followed by an afterglow broadband emission from the radio to gamma-rays, explained by a forward shock formed by the interaction of relativistic jets with the circum-burst material and reverse shock. Recent observations by Swift of X-ray flares and extended emissions introduced to this simple picture late-time internal dissipation effects.

On Aug. 27, 2017, a splendid example of multi-messenger observation by the three interferometers LIGO and VIRGO of a binary star merger event, GW170817, and the coincident observation of gamma-rays 1.7~s after it with many observations across the electromagnetic spectrum provided many fundamental physics and astronomical observations, e.g., the Hubble constant determination (see Sec.~\ref{sec3}), the identification of heavy metals in kilonova light curves, the verification of speed of GWs against the speed of light,...\cite{Ghirlanda:2018uyx}. While this observation directly connected short GRBs to kilonova, the origin of short GRBs from kilonova and long GRBs from stellar collapse is not one by one (see T.~Piran's contribution at this conference). It was also challenged by the long GRB 211211A with optical-infrared emission pointing to a binary merger or kilonova origin \cite{Mei:2022}. This observation sets the path for the synergy between GW interferometers and the future CTAO, which could be alerted by the merger observations. Observation from the ground has proved to be feasible by current ground-based IACT arrays and EAS also providing serendipitous observations. GRB 180720B, GRB 190829A and GRB 190114C above 100, 200, 300~GeV, respectively have been detected by H.E.S.S. \cite{HESS:2021dbz} and MAGIC \cite{Arakawa:2019cfc} and their measured Spectral Emission Distributions (SEDs) favor the observation of the Inverse Compton (IC) component in  SSC scenarios, where emitted synchrotron photons due to gyrating electrons in intense magnetic fields up-scatter by IC on the same emitting lepton population. More recently, GRB 221009A was initially detected by Swift and Fermi-GBM and LAT. About 2000~s later, it was then detected by LHAASO beyond 500~GeV with $\sim 100\sigma$ significance. up to about 10~TeV \cite{2022GCN.32677....1H}. These observations can be explained by IC scenarios on external seed photons e.g., from the kilonova radiation, named External Inverse Compton (EIC), or more exotic phenomena such as axion photon conversion \cite{Gonzalez:2022opy}. 

It has been noted in \cite{PhysRevD.102.083023} that, as UHECRs must be accelerated and escape before they lose their energy due to synchrotron radiation, a boundary condition to the magnetic field in the plasma reference frame can be derived for a source producing the IceCube PeV neutrinos and the UHECRs. The non-observation of neutrinos from GRBs by IceCube \cite{IceCube:2022rlk,IceCube:2017amx,IceCube:2016ipa} disfavors these powerful yet mysterious sources as accelerators of UHECR sources. In the frame of a single-zone fireball model and considering protons, the neutrino limits on the prompt phase imply that the fraction of total kinetic energy in non-thermal protons is less than 10\% or that the magnetization is so high that protons lose all their energy before producing high-energy enough neutrinos~\cite{Lucarelli:2022ush}. Nonetheless, such limit is relaxed by assuming acceleration of heavy nuclei rather than protons and/or models of multiple production zones of neutrinos and gamma-rays.

\subsection{Jetted AGNs}

An upper limit was set by IceCube on gamma-ray emitting blazars in the 2LAC catalog of Fermi~\cite{2017ApJ...835...45A}, dominated by blazars composed of Flat Spectrum Radio Quasars (FSRQs) and BL Lacertae. The $> 60$~TeV neutrino diffuse flux might be explained by these blazars up to 27\% for $E^{-2.2}$ and 50\% for $E^{-2.5}$. Despite these limits making the strong assumption of a common spectral shape of all AGNs, it is reasonable to assume that another class of sources might contribute to the diffuse neutrino flux (see Sec.~\ref{sec:calorimetric}).

A few compelling observations on blazars as neutrino emitters exist by now, possibly hinting at a particularly powerful class of blazars, which might or not have coincident flares in the $\gamma$-ray band \cite{Halzen_flare,2022arXiv220200694H}.
One of them concerns the 270~TeV neutrino alert sent by IceCube on Sep.~2017 to the astronomical community~\cite{IceCube:2018dnn}. {\it Fermi}-LAT, as well as MAGIC follow-up observations~\cite{MAGIC:2018sak}, confirmed the presence inside the error region of about $0.5^\circ$ from the IceCube event direction of a flaring blazar, TXS 0506+056, located at a redshift of 0.33 with a chance probability of 3$\sigma$.

The picture appeared immediately complex for TXS 0506+056, previously classified as high-frequency peaked BL Lacertae (HBL BL Lac) \footnote{HBL and low-frequency peaked LBL are the blazars with the frequency of the synchrotron bump of the SED  $>10^{15}$~Hz or $<10^{14}$~Hz, respectively.}, possibly being a `masquerading' FSRQ~\cite{2019MNRAS.484L.104P}.  FSRQs are relatively luminous LBL blazars with strong, optical-UV emission lines in addition to the non-thermal continuum, a signature of the Broad Line Region (BLR) outside the jet, likely photo-excited by a radiatively efficient accretion disk around a supermassive black hole (SMBH). Masquerading FSRQ would dissimulate BL Lacs as broad lines are not clearly visible due to non-thermal jet emission.
Typical leptonic models for BL Lacs are Synchrotron Self Compton models (SSC), where emitted synchrotron photons up-scatter by Inverse-Compton (IC) on the same emitting lepton population (hence named single zone models), while, for FSRQ, IC can happen on external fields provided by the thermal photons outside of the jet (External Inverse Compton models - EIC). 

\begin{wrapfigure}{L}{0.55\textwidth} 
  \centering
      \includegraphics[width=0.55\textwidth]{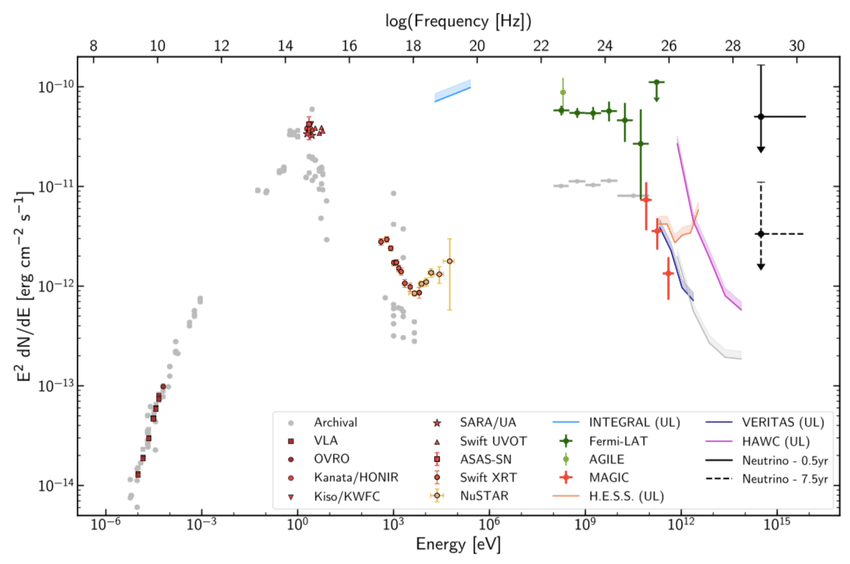} 
\caption{\label{fig:spectrum} \small \it
    SED for the blazar TXS 0506+056, based on observations obtained within 14 d of IceCube-170922A. Archival observations are in gray. The $\nu_{\mu}+\nu_{\mu}$ flux upper limits, that produce on average one detection like IceCube-170922A over a period of 0.5 (solid black error lines) and 7.5~yr (dashed black error lines) assuming an $E^{-2}$ spectrum at the most probable neutrino energy of $\sim 300$~TeV~\cite{IceCube:2018cha}.
      } 
\vspace{-10pt}
\end{wrapfigure}

The MWL SED, including the 2017 flare, cannot be explained by pure hadronic models where gamma-ray emission is dominated by proton synchrotron \cite{2001APh....15..121M} (see Fig.~\ref{fig:spectrum}). Single-zone models can barely accommodate the 2017 flare, but predict low rates of neutrinos. These can be increased ($\sim 0.1 \nu/\rm yr$) in lepto-hadronic models, but they have demanding energetics for protons the jet ($> 10^{46 - 47}$~erg/s), at times super-Eddington~\cite{2018ApJ...864...84K,2019NatAs...3...88G,Cerruti:2018tmc}. Solutions with proton synchrotron on external photon fields, e.g. on thermal radiation \cite{2018ApJ...864...84K}, radiatively inefficient accretion flow (RIAF) \cite{2019MNRAS.483L.127R} and layered jet \cite{MAGIC:2018sak} are less demanding on the energetic estimates, but do not fit the second more significant flare of $\sim 100$~d between 2014-2015 observed in the historical data of IceCube with the significance of $3.5\sigma$\cite{IceCube:2018cha}.

Single zone models do not fit the higher neutrino flux implied by this observation as the larger amount of photons from pions in the emitting region is in tension with the measured Swift X-ray flux \cite{2019ApJ...874L..29R}. This can be shifted to MeV energy and possibly imply an opaque region in photons emitting neutrinos physically separated, hence incompatible with single-zone models.

While both TXS 0506+056 flares do not pass the $5\sigma$ significance discovery threshold, some interesting observations were made in coincidence with the 2017 and 2014-2015 observations by IceCube. E.g. the MASTER global optical telescope network reported that TXS 0506+056 was in a quiet state 73~s after the IceCube 2017 event, but 2~hr after it the optical flux increased at $50\sigma$ level on top of the background, namely the biggest variation they recorded since 2005 \cite{2020ApJ...896L..19L}. Additionally, during the 100~d neutrino-flare period across 2014-2015, another longer flare with about 3.5$\sigma$ significance was observed.
VLBA 15~GHz observations in 2009-18 on the structure of the jet of TXS 0506+056, showed a curved jet structure, possibly a precessing jet with 10~yr period, with the 2017 alert in the bright precession phase or a cosmic collider of 2 jets~\cite{2019A&A...630A.103B}. 
Additionally, in Ref.~\cite{2020A&A...633L...1R} it was noted that the 43~GHz radio VLBI observations between Nov.~2017 and May~2018 indicate a compact core with highly-collimated jet and a downstream jet showing a wider opening angle (slower) external sheath. 
The slower flow could serve as seed photons for interactions producing neutrinos.
Such a scenario, also advocated in \cite{MAGIC:2018sak}, is assumed in spine-sheath models of neutrino production \cite{2014ApJ...793L..18T,2018ApJ...864...84K}.

In summary, the TXS 0506+056 evidence revealed the option that multi-messenger programs might explore the structure of jets in synergy with radio observations \cite{Petropoulou_2020}, especially in synergy with detailed radio observations. Further observations of MAGIC and VERITAS \cite{MAGIC:2022gyf} in 2017-19, resulted in another flare with 4$\sigma$ significance in Dec. 2018 detected by MAGIC in 74 hr of observations, which could have been too short to produce evidence in IceCube.


Another golden alert of IceCube, a $\nu$ event of $\sim 300$~TeV~\cite{Stein},  was compatible with the location of PKS 1502+106. This is a high-redshift ($z\sim 1.8$) LBL and highly polarized FSRQ \cite{PKS_Fermi}. No gamma-ray flare was identified but rather an increase of the radio flux measured by the 40~m telescope OVRO \cite{Kun}. This is similar to the 2014-2015 flare of TXS 0506+056
during which the radio flare was increasing, as during the 2017 event.
A similar scenario of jet precession as for TXS 0506+056 was identified in the radio data by \cite{Britzen2021}: a precessing curved jet interacting with NLR clouds at a distance of 330 pc. A ring-like and arc-like configuration develops right before the neutrino emission, not present at all times. 

It was speculated in \cite{Halzen_flare,Kun} that TXS 0506+056, PKS 1502+106, and even PKS B1424-418 (observed with one of the biggest cascade $\nu$-event ever observed by IceCube, named Big Bird) belong to a sub-class of $\sim 5\%$ blazars able to explain the IceCube diffuse flux. They could as well be efficient neutrino emitters while they are inefficient gamma-ray emitters. 
 
Additionally, a group of researchers found that another blazar PKS 0735+178 is just outside the localization error of $\sim 13^\circ$ at 90\% c.l. of the golden alert IceCube-211208A on Dec.~21, 2021 of most probable estimated energy of 172~TeV\cite{2022arXiv220405060S}. 
It was noted the coincidence that a Baikal-GVD event of most probable estimated energy of 43 TeV was detected about 4 hr after IceCube-211208A and inside the $5.5^\circ$ (50\% c.l. statistical error region)  from \cite{2021ATel15112....1D}. In \cite{2022arXiv220405060S}, it is remarked that the source doubled its X-ray flux on Dec. 17, 2021, less than $5 \times 10^3$ s of the IceCube event, and was in a very high state in $\gamma$-rays, UV and optical and other observations are quoted by Baksan and KM3NeT (and references therein).
This flaring source is of the class considered in
\cite{Plavin:2020emb,Plavin:2020mkf}. These AGNs are selected through VLBI data with strong parsec-scale cores and radio flares are searched for to exploit possible coincidence with neutrinos. A new catalog of more than 4000 sources is in \cite{Koryukova:2022abw}. While the authors have claimed a statistical excess, it is still a matter of debate worth further investigation. ANTARES has found an excess when using this catalog \cite{Illuminati:2021ezq}, which was not confirmed by IceCube.

\subsection{Calorimetric sources}
\label{sec:calorimetric}

Starburst Galaxies (SBGs) are characterized by a high formation rate (SFR), typically ranging from 10 to 1000~$M_\odot$/yr, proportional to the IR emission. Hence, SNe and star winds can be very efficient CRs factories and accelerators.
The SFR is also correlated to the radio/gamma-ray luminosity, as it is proportional to the number of acceleration sites such as supernova \cite{Kornecki:2021iuh,2021MNRAS.503.4032A}. 
These are sub-kpc regions, called Starburst Nuclei (SBNi), with gas density $>10^2$~cm$^{-3}$, IR emission $> 10^3$~eV cm$^{-3}$ and $B \gtrsim 10^2 \mu$G (indicating a high level of turbulence). They are acceleration sites of CRs, as well as the wind structures departing from them. Wind bubbles can accelerate CRs in a standard diffusive shock acceleration scenario to $\sim 100$~PeV for hundreds Myr\cite{Peretti:2021yhc}. 
In this scenario, it was found that SBG can explain the UHECR and diffuse $\nu$ flux~\cite{Condorelli:2022vfa}.

Some SBGs are identified as Syfert galaxies, making them intriguing objects where the core may host a hidden jet. While blazars shoot their jets against us, Seyfert galaxies are seen across the torus (Seyfert II), the accreted matter surrounding the SMBH, or through the narrow (NLR) and BLR (Seyfert I). Broadening is related to the Doppler shift, so to the temperature of the gas heated in the inner part by accretion emission (NRL) and for the BLR also to velocity dispersion and orbital spin around the SMBH. 


Recently, in an analysis initiated at the University of Geneva from which the filtered sample of  10 yr sample of IceCube data emerged \cite{IceCube:2022der,2021arXiv210109836I}, the Seyfert 2 starburst galaxy NGC 1068 emerged as the hottest source in a catalog of 110 gamma-ray emitters and 8 SBGs, as well as the hottest spot in the full scanned sky map search, despite of the large trial factor \cite{IceCube:2019cia}.
About 50 signal-like events were fit with a reconstructed spectrum for a single power-law hypothesis of about $E^{-3.2}$. A cumulative emission population study of the catalog produced evidence of an excess of $3.3\sigma$ with respect to the atmospheric background. The excess is dominated by NGC 1068, with a post-trial significance of $2.9\sigma$ among the background dominantly of atmospheric neutrinos. 
All other sources than NGC 1068 contributing principally to this $3.3\sigma$ excess are possble masquerading FSRQ \cite{2022arXiv220405060S}, namely TXS 0506+056, PKS 1424+240, and GB6 J1542+6129. TXS 0506+056  was expected from the former observation of the 2017 alert and second flare.   
NGC 1068 is one of the first spectroscopic AGN detection with M81 in 1909 \cite{1909PA.....17..504F}. In 1943 Seyfert observed broad line emissions from NGC 1068 and NGC 4151 \cite{1943CMWCI.671....1S}. Both Seyfert galaxies are compatible with regions with event excesses in an updated IceCube analysis \cite{IceCube:2022der}. For NGC 1068 about 79 signal-like events are fit with a reconstructed spectrum for a single power-law hypothesis of about $E^{-3.2}$ providing evidence at $4.5\sigma$ c.l. that neutrinos are messengers from this well-known source at only 14.4~Mpc from us. Nonetheless, the inferred neutrino flux is higher by about an order of magnitude than the MAGIC upper limits \cite{2019ApJ...883..135A} and the Fermi-LAT fluxes up to about 10 GeV \cite{Fermi-LAT:2012nqz}. 

The absence of the gamma-ray counterpart to neutrinos in the TeV region triggered AGN corona models \cite{Inoue_2021,2022arXiv221104460M}. NGC 1068 is a composite system and there might be the contribution of various acceleration processes: it hosts a highly obscured mildly relativistic jet seen in the radio through its accretion disk (as it is a Seyfert 2 galaxy) interacting with interstellar matter or a molecular cloud \cite{1996ApJ...458..136G} and eventually originating gamma-rays from IC on IR radiation in the starburst region \cite{2011A&A...535A..19L}. Two zone models where gamma-ray emission above 1~GeV results predominantly from the starburst region and TeV neutrinos in the corona have been proposed \cite{Eichmann:2022lxh}. Very hard spectra compatible with the acceleration of CRs have been obtained in AGN-driven wind models in the circum-nuclear molecular disk \cite{Lamastra:2019zss}. Shock acceleration might take place also in the starburst region, in particular in wind bubbles emerging from the observed radio starburst nuclei with consequent proton-proton interactions \cite{Peretti:2021yhc}. Such a composite object will be an interesting target for CTAO to explore the interplay between the AGN and starburst nature.

\section{The granted flux of neutrinos from the Galactic plane}
\label{sec:grantedfluxes}


At energy below the discovered diffuse neutrino flux, NTs are beginning to observe the granted diffuse neutrino flux produced by CRs interacting with the interstellar matter in the Galaxy. 
Recently, ANTARES published the observation of an excess incompatible with the atmospheric neutrino background at $2\sigma$ c.l. with a preferred spectrum of $E^{-2.45}$ from the Galactic Ridge (galactic longitude $|\ell|<30^\circ$ and galactic latitude $|b|<2^\circ$)~\cite{ANTARES:2022izu}. 
Despite the large error, the measured neutrino flux is compatible with a neutral pion decay model from interactions of protons with a power-law spectrum of $E^{-2.4}$, harder than the local CR spectrum of $E^{-2.7}$~\cite{ParticleDataGroup:2022pth}. 
The model is normalised to the gamma-ray flux measured by the Fermi-LAT from the same Galactic Ridge region between 19 GeV and 3~TeV~\cite{Neronov:2019ncc}. It has been speculated that such a hard spectrum could include the contribution of another component, such as decays of heavy dark matter in the Galactic ridge~\cite{Neronov:2018ibl}. Previous results are upper limits to the model in Ref.~\cite{2015ApJ...815L..25G} from IceCube only~\cite{2017ApJ...849...67A} and with ANTARES data~\cite{ANTARES:2018nyb}, which already indicated that the flux will be soon observed and that its contribution to the diffuse IceCube neutrino flux at $> 60$~TeV is $<10\%$~\cite{Ahlers:2015moa}.
As already noted in Sec.~\ref{sec:detectors}, new analyses exploiting cascade-like events dominated by electron neutrinos will push the energy threshold down increasing sensitivity to the galactic plane flux, as the atmospheric electron neutrino background flux is one order of magnitude lower than the muon one from pion and kaon decays.  

The ultra-high energy gamma-ray measurements (e.g. from Tibet AS$\gamma$ \cite{TibetASgamma:2021tpz} and future expected ones by LHAASO) are relevant as the CR proton spectrum is still not perfectly known in the region of the knee~\cite{2022A&A...661A..72B,PhysRevD.98.043003}. For instance, there is some discrepancy between KASCADE and IceTop proton knees at about 4~PeV, and to some extent to the discrepancies in the helium knee at about $Z \times 4$~PeV, where Z is the number of protons in the ionized nucleus~\cite{PhysRevD.98.043003}. 


New results by LHAASO~\cite{Luque:2022buq,Zhao:2021dqj} and Tibet As$\gamma$~\cite{TibetASgamma:2021tpz} extend to ultra-high energies the measurement of the flux from the Galactic Plane up to PeV energies, indicating the presence of PeVatron accelerators in the Galaxy, still unknown as standard Diffuse Shock Acceleration  applied to supernova shocks only achieves one order of magnitude lower energies. 
IceCube and the KM3NeT upcoming neutrino telescope in the Mediterranean Sea in synergy with CTAO, LHAASO and Tibet As have the potential to unravel the origin of the knee \cite{PhysRevD.98.043003,2022A&A...661A..72B,Ahlers:2015moa,Luque:2022buq}, to understand its propagation or acceleration origin, explore diffusion processes in the Galaxy and dark matter in the Galactic Plane.

\section{Conclusions}
In summary, I discussed the field of gamma-ray astrophysics and that there are a few solid observations on potential neutrino sources, and many hints, that need multi-messenger and multi-observatory observations.

\subsection{Acknowledgments}
I wish to thank the organisers of this very high-level conference, and particularly Prof. Felix Aharonian and Prof. Werner Hofmann for the Gamma recurrent appointment.

\bibliographystyle{iopart-num}
\bibliography{multimessenger}

\end{document}